\documentclass[aps,prd,amsmath,amssymb,10pt,letterpaper,balancelastpage,showpacs,superscriptaddress,nofootinbib,floatfix,notitlepage,longbibliography
]{revtex4-2}
\usepackage[utf8]{inputenc}
\usepackage{bm}
\usepackage[utf8]{inputenc}
\usepackage{bm}

\newcommand{\pbh}{\rm MACHO}

\usepackage[dvipsnames]{xcolor}
\definecolor{red}{rgb}{0.9, 0,0}
\definecolor{cerulean}{rgb}{0., 0.42,0.9}
\definecolor{navy}{rgb}{0.05, 0.05,0.8}

\usepackage[colorlinks]{hyperref}
\hypersetup{
  colorlinks = true,
  citecolor = red,
	linkcolor = navy
}
\usepackage{comment}
\usepackage{caption}
\usepackage{subcaption}
\usepackage{bbm}
\usepackage{amsfonts}
\usepackage{mathrsfs}
\usepackage{bbold}
\usepackage{amsmath,amssymb}
\usepackage{subcaption}
\usepackage{siunitx}
\usepackage{slashed}

\usepackage{epsfig}
\usepackage{graphicx}   
\usepackage{url}
\usepackage{float}
\usepackage{soul}
\usepackage{pstricks}
\usepackage{color}
\usepackage{multirow}

\begin{document}

\title{Constraints on Dark Matter from Dynamical Heating of Stars in Ultrafaint Dwarfs. Part 1: MACHOs and Primordial Black Holes}

\author{Peter W.~Graham}
\affiliation{Stanford Institute for Theoretical Physics, Department of Physics, Stanford University, Stanford, CA 94305, USA}
\affiliation{Kavli Institute for Particle Astrophysics \& Cosmology, Department of Physics, Stanford University, Stanford, CA 94305, USA}

\author{Harikrishnan Ramani}
\affiliation{Stanford Institute for Theoretical Physics, Department of Physics, Stanford University, Stanford, CA 94305, USA}

\begin{abstract}
We place limits on dark matter made up of compact objects significantly heavier than a solar mass, such as MACHOs or primordial black holes (PBHs).  In galaxies, the gas of such objects is generally hotter than the gas of stars and will thus heat the gas of stars even through purely gravitational interactions.  Ultrafaint dwarf galaxies (UFDs) maximize this effect.  Observations of the half-light radius in UFDs thus place limits on MACHO dark matter.  We build upon previous constraints with an improved heating rate calculation including both direct and tidal heating, and consideration of the heavier mass range above $10^4 \, M_\odot$.  Additionally we find that MACHOs may lose energy and migrate in to the center of the UFD, increasing the heat transfer to the stars.  UFDs can constrain MACHO dark matter with masses between about $10 M_\odot$ and $10^8 M_\odot$ and these are the strongest constraints over most of this range.
\end{abstract}

\maketitle

\tableofcontents

\section{Introduction}

The particle identity of dark matter (DM) continues to be one of the intriguing mysteries of particle physics today. While a new elementary particle would have mass below the Planck mass, dark matter could instead consist of primordial black holes (PBHs) or composite compact objects with masses that far exceed the Planck mass. We refer to all such heavy compact dark matter objects as Massive Compact Halo Objects or MACHOs in this paper (of which PBH's would be one example).  The search for MACHOs thus probes a large part of dark matter parameter space over many orders of magnitude in mass.

MACHOs are a prediction of many Beyond-The-Standard Model (BSM) scenarios. PBHs can form as a consequence of a very large primordial power spectrum at small length scales arising from features in the inflationary potential, early universe phase transitions or collapse of topological defects. See Ref.~\cite{Green:2020jor} for a comprehensive review of formation mechanisms. Other well motivated compact objects present in BSM theories include axion stars~\cite{Kolb:1993zz}, Q-balls~\cite{coleman1985q}, non-topological solitons~\cite{lee1992nontopological}, quark nuggets~\cite{Witten:1984rs}, asymmetric DM nuggets~\cite{Wise:2014jva}, dark blobs~\cite{Grabowska:2018lnd} and mirror stars~\cite{Curtin:2019ngc}. 

In most of these scenarios, MACHOs possess only gravitational interactions with the SM. Furthermore, their large mass and hence small number density imply that astrophysics is often the only probe of this scenario. MACHOs making up 100\% of DM is heavily constrained for MACHO masses $M_{\pbh}$ greater than $10^{-10} M_{\odot}$. In the $M_{\pbh}>1M_{\odot}$ region of interest in this paper, the leading constraints arise from gravitational lensing~\cite{Macho:2000nvd,Niikura:2019kqi,EROS-2:2006ryy,Oguri:2017ock,Wilkinson:2001vv} and dynamical heating of SM astrophysical objects~\cite{Yoo:2003fr,Ramirez:2022mys, Carr:2020gox,moore1993upper}. There are also PBH specific limits that arise from gravitational waves~\cite{Kavanagh:2018ggo,LIGOScientific:2019kan} and accretion~\cite{accretion1,accretion2,accretion3,accretion4}. See also Ref.~\cite{Bai:2020jfm} which derives accretion limits on MACHOs that make up 100\% of DM. However this does not preclude the possibility that a small but non-trivial amount of DM is in MACHOs. Many of the formation mechanisms listed above, can predict DM fractions smaller than 100\% and studying this subcomponent might be the only probe of such BSM scenarios. 

In this work, we systematically study an important probe of MACHOs above $\approx 10 \, M_\odot$ mass.
If the MACHOs and the stars in a galaxy have roughly similar speeds, then the gas of such heavy MACHOs will have an effective temperature higher than that of the gas of stars.  The tendency to approach thermal equilibrium then ensures that the MACHOs must transfer energy to the stars locally at every point in the galaxy.  This `heating' will cause the stellar gas to expand\footnote{In fact, due to the negative heat capacity of such a gravitationally bound system the stars' effective temperature (average kinetic energy) will actually decrease due to this `heating' (making them effectively colder) but we will still refer to this as heating throughout this paper.}.  This effect will occur even if the only coupling between MACHOs and the Standard Model is gravitational.  For most objects in our universe the gravitational heat transfer time scale would be far too long to be observable.  However there is one known class of objects for which the gravitational heat transfer rate can be fast enough to be relevant: ultra-faint dwarf galaxies (UFDs). The cores of these UFDs are known to contain high dark matter densities $\approx 1 \, M_\odot \, \textrm{pc}^{-3}$ and low stellar and dark matter velocity dispersion compared to other dark matter environments.  Further UFDs are generally old with a tight (`cold') cluster of stars in the center. As a result they are uniquely suited to study the gravitational interactions that cause heat exchange between dark matter and stars which peak at small momentum transfer. This gravitational heating of stars accumulated over the age of the UFD causes the half-light radius of stars to increase. Thus, the observed half-light radius puts an upper bound on such heating and hence on the MACHOs in the UFD core. 
While we restrict ourselves to studying MACHOs in this work, we discuss UFD constraints on more diffuse substructure in a companion paper~\cite{pwgforthcoming}.  

Previous papers have explored stellar heating to set limits on primordial black holes~\cite{Brandt:2016aco,zhu2018primordial,Stegmann:2019wyz,zoutendijk2020muse, Wadekar:2022ymq}. In this work, we build on these previous papers with several improvements to the bounds. First, we make some corrections to the heat transfer rate equation. Second, we extend the reach to MACHO masses higher than $10^4 M_\odot$, calculating the  cutoff of the effect at higher masses. Further, in this mass range, we point out that the dynamical friction induced cooling of a sub-component MACHO population by the majority smooth DM causes the MACHOs to migrate inward in the UFD. This results in large MACHO densities near the stellar core, resulting in higher heat transfer to the stars than previously believed, which significantly improves the limits on MACHOs in the $\approx 10^4 M_\odot$ mass range. 

The rest of this paper is organized as follows. In Section.~\ref{sec:heattransfer} we describe the local heat transfer rate between two different gases caused by gravitational scattering. This will be useful for both stellar heating due to MACHOs as well as MACHO cooling due to the smooth DM. In Section.~\ref{sec:UFD}, we provide a brief overview of UFDs as well as its properties such as the ambient DM density and velocity dispersion. Section.~\ref{sec:exphlr} deals with the expansion of the stellar scale radius and considers the different kinds of heat transfer between MACHOs and stars. In Section.~\ref{sec:migration} we present the density enhancement of MACHOs near the stellar core due to heat transfer with the smooth DM component and present our final exclusion curves. Concluding remarks are presented in Section.~\ref{sec:conclusion}.

\section{Heat Transfer}
\label{sec:heattransfer}
In this section we discuss the heat transfer between two populations caused by gravitational scattering. We will apply the formalism developed here for scattering between stars and MACHOs in Section.~\ref{sec:exphlr} and MACHOs and the smooth DM population in Section.~\ref{sec:migration}. Locally each component can be treated as a gas of particles at a given temperature and density, although the temperature and density will vary over the galaxy.  We can then calculate a local (instantaneous) heat transfer rate between the two components and integrate over the galaxy.

\subsection{General heat transfer rate}

We will discuss the general heat transfer rate between two gases of particles caused solely by gravitational scattering.  Take the two gases to be $A$ and $B$ with temperatures $T_A$ and $T_B$ and mass densities $\rho_A$ and $\rho_B$ whose individual particles have masses $m_A$ and $m_B$.
We will treat each component $i$ as a gas of particles with a locally Maxwellian velocity distribution
\begin{align}
\label{eqn:maxwelldistribution}
    f_i (v) = \frac{1}{\left( 2 \pi \sigma_i^2 \right)^\frac{3}{2}} \, e^{- \frac{v^2}{2 \sigma_i^2}}
\end{align}
where $\sigma_i$ is the velocity dispersion of component $i \in \{ A, B \}$.  The temperature is $T_i = m_i \sigma_i^2$.
Let us take $A$ to be the hotter gas $T_A > T_B$ (so generally $A$ will be the MACHOs and $B$ will be the stars in Section.~\ref{sec:exphlr} and the smooth DM component in Section.~\ref{sec:migration}).
Then the local rate of heat transfer from $A$ to $B$ per unit mass in $B$ is given by~\cite{Munoz:2015bca,Dvorkin:2013cea}
\begin{align}
\label{eqn:initialheatingrateeqn}
    H_B=\frac{2 \rho_A \sigma_0 \left(T_A-T_B\right)}{\left(m_A+m_B\right)^2 \sqrt{2\pi} \, v_{\rm th}^3}
\end{align}
where $H_B \equiv \frac{1}{\rho_B} \frac{du_B}{dt}$ and $\frac{du_B}{dt}$ is the rate of change of the total energy density of the $B$ gas (including both kinetic and gravitational potential energy).
Here $v_{\rm th}\equiv \sqrt{\frac{T_A}{m_A}+\frac{T_B}{m_B}}$ and the cross-section $\sigma_0$ is defined via $\sigma_T=\frac{\sigma_0}{v_{\rm rel}^4}$, where $\sigma_T$ is the transfer cross-section for a collision between an $A$ and a $B$ particle at relative velocity $v_{\rm rel}$. Equivalently we can write the net heat transfer rate between $A$ and $B$ per unit volume as
\begin{align}
    - \frac{d u_A}{dt} = \frac{d u_B}{dt} = \frac{2 \rho_A \rho_B \sigma_0 \left(T_A-T_B\right)}{\left(m_A+m_B\right)^2 \sqrt{2\pi} \, v_{\rm th}^3}
\end{align}
Let us calculate $\sigma_T$ next.

\subsection{Transfer cross-section}

We want to calculate the transfer cross-section for two particles $A$ and $B$ undergoing purely gravitational scattering.  The gravitational Rutherford cross-section is given by
\begin{align}
    \frac{d\sigma}{d\cos \theta}=\frac{\pi}{2}G^2m_A^2 m_B^2 \frac{1}{\left(\frac{1}{2}\mu_{AB} v_{\rm rel}^2\right)^2}\frac{1}{\left(1-\cos \theta\right)^2}
\end{align}
where $\mu_{AB} = \frac{m_A m_B}{m_A+m_B}$ and $v_{\rm rel}$ is the relative velocity between the particles.
The total momentum transfer cross-section is given by
\begin{align}
    \sigma_T = \int_{\cos \left(\theta_{\rm min} \right)}^{\cos \left( \theta_{\rm max} \right) } \frac{d\sigma}{d\cos \theta}\left(1-\cos \theta\right)d\cos \theta
\end{align}
The limits of integration are fixed by the corresponding minimum and maximum impact parameters $b_{\rm min}$ and $b_{\rm max}$ via
\begin{align}
\cos \theta=\frac{b^2 \mu_{AB}^2 v_{\rm rel}^4-\alpha_{AB}^2}{b^2 \mu_{AB}^2 v_{\rm rel}^4+\alpha_{AB}^2}
\end{align}
where $\alpha_{AB}=G m_A m_B$. 
This gives,
\begin{align}
\sigma_T=\frac{2 \pi  \alpha_{AB}^2}{\mu_{AB}^2 v_{\rm rel}^4}
\log \left(\frac{\alpha_{AB}^2+b_{\rm max}^2 \mu_{AB}^2
   v_{\rm rel}^4}{\alpha_{AB}^2+b_{\rm min}^2 \mu_{AB}^2
   v_{\rm rel}^4}
   \right)
\end{align}

We can now use these expressions to find the heating rate.  We will substitute $\sigma_0=\frac{2\pi \alpha_{AB}^2}{\mu_{AB}^2} \log \left(\frac{\alpha_{AB}^2+b_{\rm max}^2 \mu_{AB}^2
   v_{\rm rel}^4}{\alpha_{AB}^2+b_{\rm min}^2 \mu_{AB}^2
   v_{\rm rel}^4}
   \right)$ and $v_{\rm th}\equiv \sqrt{\frac{T_A}{m_A}+\frac{T_B}{m_B}}$, with the temperatures defined as $T_{A,B}=m_{A,B} \, \sigma_{A,B}^2$ into Eqn~\eqref{eqn:initialheatingrateeqn}.
This gives the heating rate of $A$ per unit mass to be
\begin{align}
    H_A = - \frac{\rho_B}{\rho_A} H_B 
    &=2\sqrt{2\pi} \frac{G^2  \rho_B \left(m_B\sigma_B^2-m_A \sigma_A^2\right)}{\left(\sigma_A^2+\sigma_B^2\right)^\frac{3}{2}}\log \left(\frac{\alpha_{AB}^2+b_{\rm max}^2 \mu_{AB}^2
   v_{\rm rel}^4}{\alpha_{AB}^2+b_{\rm min}^2 \mu_{AB}^2
   v_{\rm rel}^4}
   \right) \nonumber \\
   &=4\sqrt{2\pi} \frac{G^2  \rho_B \left(m_B\sigma_B^2-m_A \sigma_A^2\right)}{\left(\sigma_A^2+\sigma_B^2\right)^\frac{3}{2}}\log \Lambda 
    \label{heateqn}
\end{align}
Here $\Lambda^2=\frac{\alpha_{AB}^2+b_{\rm max}^2 \mu_{AB}^2
   v_{\rm rel}^4}{\alpha_{AB}^2+b_{\rm min}^2 \mu_{AB}^2
   v_{\rm rel}^4}$.
Eqn~\eqref{heateqn} is the main result we use to calculate heating.

\subsection{Relation to Dynamical Friction}

Up until this point we have calculated the heat transfer between gases $A$ and $B$ from the net integrated effect of all the individual gravitational scatterings between pairs of particles.  We will now comment on the relation between this and the well-known dynamical friction formula.
If $A$ is hotter than $B$, then an $A$ particle will slow down as it is passing through the $B$ gas.  The dynamical friction formula for the slowdown of individual $A$ particles in a bath of $B$ is
\begin{align}
    \frac{d \vec{v}_A}{dt}=-\frac{4\pi G^2 (m_A + m_B) \rho_B \ln\Lambda}{v_A^3} \left[\textrm{erf}(X)-\frac{2X}{\sqrt{\pi}}e^{-X^2}\right] \vec{v}_A 
\end{align}
where $X=\frac{v_A}{\sqrt{2} \, \sigma_B}$.
The rate of change of the energy of an $A$ particle per unit mass (the `heating' rate of $A$) is $H_A = v_A\frac{dv_A}{dt}$.
Integrating this over the Maxwell Boltzmann distribution Eqn.~\eqref{eqn:maxwelldistribution} for $v_A$ yields the average heating rate of the $A$ gas per unit mass
\begin{align}
H_A= \int_0^\infty   v_A\frac{dv_A}{dt}f(v_A) 4 \pi v_A^2 dv_A
\end{align}
This gives
\begin{align}
\label{eqn:dynfricheating}
H_A=-4\sqrt{2\pi} \frac{G^2 \rho_B \left(m_A + m_B \right)  \sigma_A^2 
 }{\left(\sigma_A^2+\sigma_B^2\right)^\frac{3}{2}} \log \Lambda 
\end{align}
In the special case of $m_A\gg m_B$ this reproduces Eqn.~\eqref{heateqn}.

But clearly if we are not in the limit $m_A\gg m_B$ then the dynamical friction formula does not give the same cooling rate as Eqn.~\eqref{heateqn}, even when $A$ is hotter than $B$, namely $m_A \sigma_A^2 \gg m_B \sigma_B^2$.  We  can see that the dynamical friction formula does not give the correct answer in this case and that in fact Eqn.~\eqref{heateqn} is the fully correct answer in the general case.  This is because the dynamical friction formula only uses the first-order diffusion coefficient.  Including both first- and second-order diffusion coefficients the average heating rate per unit mass of an $A$ particle is (see \cite{binney2011galactic})
\begin{align}
\label{eqn:diffcoefheating}
    H_A= \int_0^\infty  \left( v_A D[\Delta v_\parallel] + \frac{1}{2} D[\Delta v_\parallel^2 ] + \frac{1}{2} D[\Delta v_\perp^2 ] \right) f(v_A) 4 \pi v_A^2 dv_A
\end{align}
where the first-order diffusion coefficient is
\begin{align}
    D[\Delta v_\parallel] = -\frac{4\pi G^2  \rho_B (m_A + m_B) \ln\Lambda}{\sigma_B^2} F[X]
\end{align}
and the second order diffusion coefficients are
\begin{align}
    D[\Delta v_\parallel^2 ] &= \frac{4 \sqrt{2} \pi G^2 \rho_B m_B \ln\Lambda}{\sigma_B} \frac{F(X)}{X} \\
    D[\Delta v_\perp^2 ] &= \frac{4 \sqrt{2} \pi G^2 \rho_B m_B \ln\Lambda}{\sigma_B} \left( \frac{\textrm{erf}(X) - F(X)}{X} \right)
\end{align}
where $X=\frac{v_A}{\sqrt{2} \, \sigma_B}$ as above and
\begin{align}
    F[X] = \frac{1}{2 X^2} \left( \textrm{erf}(X)-\frac{2X}{\sqrt{\pi}}e^{-X^2} \right)
\end{align}
Using these in Eqn.~\eqref{eqn:diffcoefheating} gives
\begin{align}
\label{eqn:heateqn2}
    H_A = 4\sqrt{2\pi} \frac{G^2  \rho_B \left(m_B\sigma_B^2-m_A \sigma_A^2\right)}{\left(\sigma_A^2+\sigma_B^2\right)^\frac{3}{2}}\log \Lambda 
\end{align}
This is the same as Eqn.~\eqref{heateqn}.  Thus dynamical friction Eqn.~\eqref{eqn:dynfricheating} is not the general expression and is only correct in the limit $m_A \gg m_B$.  When we are not in this limit then we see that dynamical friction generally overestimates the cooling rate.  When $A$ is not much heavier than $B$, the $A$ particles feel significant diffusive forces through their collisions with $B$.  This causes some random walk of the $A$ velocity, rather than just friction, which partially cancels the cooling power of the dynamical friction.

Eqn.~\eqref{heateqn} (equivalently Eqn.~\eqref{eqn:heateqn2}) is our main result for the heating rate. Note that this formula differs  from more approximate formulas used in similar situations (see e.g.~\cite{Brandt:2016aco,Church:2018sro,lacey1985massive}) by $\mathcal{O}(1)$ numerical factors as well as some parametric differences (e.g.~in the Coulomb log).  Some of the reasons for the differences are: we have integrated over a distribution of velocities for the MACHOs instead of taking them all to be at the same speed, we have included all the diffusion coefficients, and we have taken the full expression for the Coulomb log instead of just using the ratio of impact parameters.

So ultimately Eqn.~\eqref{heateqn} gives the general expression for the heating rate per unit mass.  Identifying $A$ with MACHOs and $B$ with stars, it is clear that to maximize heat transfer, we need an environment with high DM density and low velocity dispersion.   We turn our attention to this next.

\section{Ultra-faint dwarfs} 
\label{sec:UFD}
Ultra-faint dwarf galaxies are excellent objects to use for constraining this heating effect due to their high DM densities, low velocity dispersions, old ages, and relatively small stellar scale radii.
We use the candidate ultra-faint dwarf Segue-I. Its 2D projected half-light radius $r_{h}= 24.2~\textrm{pc}$\cite{munoz2018megacam} and the luminosity weighted average of the line-of-sight velocity dispersion $\sqrt{\langle\sigma^2_{\star \rm los}\rangle}=3.7 \, \textrm{km~s}^{-1}$\cite{simon2011complete}. The dark matter density and velocity dispersion are inferred based on an assumed model. We will next lay forth the assumptions and then estimate these quantities. 

The stars are assumed to follow a Plummer sphere
\begin{align}
    \rho_\star(r) = \frac{3M_\star}{4\pi R_{0,\star}^3}\left(1+\frac{r}{R_{0,\star}}\right)^{-\frac{5}{2}}
\end{align}
where the stellar scale radius $R_{0,\star}$ is related to the 
2D projected half-light radius $r_h$ via $R_{0,\star}=\frac{4}{3} \sqrt{2^{2/3}-1} \, r_h\approx r_h$ and $M_\star\approx 10^3 M_{\odot}$ is the total stellar mass. $R_{0,\star}=24.73~\textrm{pc}$ since $r_h=24.2~\textrm{pc}$. 

The DM is assumed to be distributed in a Dehnen sphere, with mass density
\begin{align}
    \rho_{\rm DM}(r)=\frac{(3-\gamma) M_{\rm UFD}}{4\pi R_{\rm UFD}^3}\left(\frac{r}{R_{\rm UFD}}\right)^{-\gamma}\left(1+\frac{r}{R_{\rm UFD}}\right)^{\gamma-4}
\end{align}
where $R_{\rm UFD}$ is the dark matter scale radius and $M_{\rm UFD}$ is the total dark matter mass. We take $M_{\rm UFD}=10^9 M_\odot$ in accordance with N-body simulations that predict the halo mass - stellar mass relationship~\cite{garrison2014elvis,zhu2018primordial}.\footnote{Other simulations predict even smaller halo masses~\cite{read2016dark} for dwarfs, which reduce $\sigma_{\pbh}$ thus increasing the heat transfer between DM and stars, which ultimately means this is a conservative choice for us.}

Let us now relate the observables to DM parameters. 
The luminosity weighted average of the line-of-sight velocity dispersion is given by~\cite{Wolf:2009tu},

\begin{align}
    \langle \sigma^2_{\star \rm{los}}\rangle&=\frac{G M_\frac{1}{2} }{3R_{\frac{1}{2},\star}}
    \label{losvel}
\end{align}
Here, $M_\frac{1}{2}$ is the mass contained inside the de-projected half-light radius $R_{\frac{1}{2},\star}$. This quantity, not to be confused with the 2D projected half-light radius, is related to it via, $R_{\frac{1}{2},\star}=\frac{4}{3}r_h$

We conservatively choose the cored density profile for the dark matter i.e. $\gamma=0$. \footnote{This is a conservative choice, afterall, a cuspy profile such as $\gamma=1$ or NFW only increases the density of dark matter in the core, further increasing the heating effect on the stars.} Then, Eqn.~\eqref{losvel} can be rewritten as,  
\begin{align}
 \langle \sigma^2_{\star \rm{los}}\rangle|_{\gamma=0}&=\frac{G M_{\rm UFD}}{3R_{\rm UFD}^3} R^2_{\frac{1}{2},\star}
\end{align}
Here we have assumed that the dark matter mass density in the core is significantly higher than the stellar mass density so that we can use only the total dark matter mass $M_{\rm UFD}$.  Current observations of the UFD justify this assumption.
We thus obtain $R_{\rm UFD}\approx 446.3~\textrm{pc}$. 
The DM density, well inside the core is roughly 
\begin{align}
    \rho_{\rm DM}|_{\rm core}\approx 2.46 M_\odot~\textrm{pc}^{-3}
\end{align}.

For later use, we calculate the velocity dispersion of DM and the stellar population at the stellar scale radius. These are determined by invoking hydrostatic equilibrium~\cite{binney1980radius}.
In the $R_{0,\star} \ll  R_{\rm UFD}$ limit, we can calculate the velocity dispersion at the stellar scale radius $R_{0,\star}$ to be
\begin{align}
\sigma^2_\star\left(r = R_{0,\star}\right)&\approx\frac{2 G M_{\rm UFD} R_{0,\star}^2}{3 R_{\rm UFD}^3}+\frac{G M_\star}{6 \sqrt{2} R_{0,\star}}
    \label{velstar}
\end{align}
Here we have {\it not} assumed that the dark matter mass dominates the stellar mass inside $R_{0,\star}$.  While dark matter does dominate today, we will also use this formula at earlier times when it is possible that the stars were more dense than they are today.  Then we can evaluate the stellar velocity dispersion at the half-light radius today as

\begin{align}
\sigma_\star \left(R_{0,\star}=24.73 \, \textrm{pc}\right)
    \approx 4.45 \, \textrm{km}/{\rm sec}
\end{align}

Next, we assume that some or all of the DM in the UFD is made up of MACHOs of the same mass $M_{\pbh}$, and their density is $\rho_{\pbh}=f_{\pbh} \rho_{\rm DM}$, i.e. MACHOs make up a mass fraction $f_{\pbh}$ of all of DM. The rest of DM is assumed to be a very light particle in a smooth profile with velocity dispersion $\sigma_{\rm smooth}$. We also assume that the MACHOs have the same density profile and hence velocity dispersion as the majority DM, i.e. $\sigma_{\rm DM}\equiv\sigma_{\rm smooth}=\sigma_{\pbh}$. 

This velocity dispersion at $r=R_{0,\star}$ can be obtained by solving the hydrostatic equilibrium condition to give,
\begin{align}
   \sigma^2_{\rm DM}=\sigma^2_{\rm smooth}=\sigma^2_{\pbh}(r=R_{0,\star})&\approx G \frac{M_\star}{\sqrt{2}R_{0,\star}} \nonumber \\
     &+\frac{G}{R_{\rm UFD}}\left(\frac{M_{\rm UFD}}{30}+M_\star\left\{9.08+4\log\frac{R_{0,\star}}{R_{\rm UFD}}\right\}\right) \nonumber  \\
     &\approx \frac{G}{R_{\rm UFD}}\frac{M_{\rm UFD}}{30}\nonumber  \\ 
     &\approx \left(18.0 \textrm{km}/{\rm sec}\right)^2
     \label{veldm}
\end{align}
In the penultimate line we have used that the dark matter potential term always dominates the stellar potential term even at our most conservative initial conditions when $R_{0,\star}$ is at its smallest.

\section{Expansion of stellar scale radius}
\label{sec:exphlr}
Next, we study the effect of heat transfer from $\pbh$s into the star system. We follow ~\cite{Brandt:2016aco} and make a simplifying assumption that under heating, the star system expands while retaining its shape, i.e. we model the effect of heating by an increase in stellar scale radius $R_{0,\star}$ while the stars remain in a Plummer profile. 

To quantify the increase in $R_{0,\star}$, we need to equate the heat transfer from $\pbh$s to stars, to the rate of change of total energy of stars, which should depend on $R_{0,\star}$.

\begin{align}
     \frac{d\mathcal{E}_{\star}}{dt}= \int dr \, 4\pi r^2  \rho_\star(r,R_{0,\star}) \, H_\star \left( \sigma_\star(r,R_{0,\star}),\sigma_{\pbh}(r,R_{0,\star}),\rho_{\rm \pbh}(r,R_{0,\star}) \right) 
    \label{heatint}
\end{align}
where $H_\star$ is defined as the heating rate of stars per unit mass, see Eqn.~\eqref{heateqn} and $\mathcal{E}_\star$ is the total energy in the stellar system.
However in Section.~\ref{sec:UFD}, it was shown that for the specific choice of Plummer profile for the stars and Dehnen profile for the DM,  $\sigma_{\pbh}$ and $\rho_{\rm \pbh}$ are $r$ independent and  $\sigma_\star \ll \sigma_{\pbh}$ and can be ignored. Hence, the integral in Eqn.~\eqref{heatint} can be done trivially to obtain,
\begin{align}
    \frac{d\mathcal{E}_{\star}}{dt}= M_\star H_\star(R_{0,\star}) 
    \label{dedtleft}
\end{align}

The subtlety of identifying the relevant total energy in stars, $\mathcal{E}_\star$ when both the heat source ($\pbh$s) and heat sink (stars) are in the same gravitational system is treated in detail in Appendix.~\ref{totenergy}. We find that, in the $f_{\pbh}\ll 1$ limit, we can express the LHS in terms of the stellar scale radius, $R_{0,\star}$, and the other relevant quantities (see Appendix.~\ref{totenergy}) and find:
\begin{align}
\frac{d\mathcal{E}_\star}{dt}= G M_\star\frac{dR_{0,\star}}{dt}
\times \left\{\frac{3 \pi  M_\star}{64 R_{0,\star}^2}+\frac{M_{\rm UFD} R_{0,\star}}{4R_{\rm UFD}^3}\left(6  \log
   \frac{R_{\rm UFD}}{R_{0,\star}}+  (3 \log (4)-17)\right)\right\}
   \label{dedt}
\end{align}
While this was shown to be true only in the $f_{\pbh}\ll 1$ limit in Appendix.~\ref{totenergy}, we extrapolate it all the way to $f_{\pbh}=1$. We expect this to be at most $\mathcal{O}(1)$ discrepant, with our limits being conservative. 

Substituting Eqn.~\eqref{dedt} into the LHS of Eqn.~\eqref{dedtleft}, we get,
\begin{align}
    \frac{dR_{0,\star}}{dt}=H_\star(R_{0,\star})\times\left\{\frac{3 \pi  G M_\star}{64 R_{0,\star}^2}+\frac{G M_{\rm UFD} R_{0,\star}}{4R_{\rm UFD}^3}\left(6  \log
   \frac{R_{\rm UFD}}{R_{0,\star}}+  (3 \log (4)-17)\right)\right\}^{-1}
    \label{heatingeqndiff}
\end{align}
where we have assumed that the only result of the heating is that $R_{0,\star}$ changes in time (but otherwise both the stellar and DM profiles remain the same).

We next evaluate the heating rate in different regimes

\subsection{Direct Heating}
\label{secdirectheating}
We start by estimating the heating rate for the smallest mass MACHOs. In this regime the density of MACHOs is high enough that  over the age of the UFD some MACHOs pass within the scale radius of the stellar distribution.  We thus conservatively include only the heating from these passes and not from MACHOs further out, and call this the `direct heating' regime.
In the next section we will calculate the tidal heating from MACHOs passing well outside the stellar distribution.  Our ultimate limit will come from adding these two heating rates.

This heating rate is given by
Eqn.~\eqref{heateqn}. We identify species $A$ in Eqn.~\eqref{heateqn} with stars and species $B$ with the MACHOs. Thus 

\begin{align}
    H_{\star,\rm direct}
    =2\sqrt{2\pi} G^2  \rho_{\pbh} \frac{ \left(M_{\pbh}\sigma_{\pbh}^2-m_\star \sigma_\star^2\right)}{\left(\sigma_\star^2+\sigma_{\pbh}^2\right)^\frac{3}{2}}&\log \left(\frac{\alpha^2+b_{\rm max}^2 \mu^2
   v_{\rm rel}^4}{\alpha^2+b_{\rm min}^2 \mu^2
   v_{\rm rel}^4}
   \right)\nonumber \\
 = ... &\log \left(\frac{b_{90}^2+b_{\rm max}^2}{b_{90}^2+b_{\rm min}^2 }\right)
    \label{eqn:heateqnspec}
\end{align}
Here, $\alpha=G m_\star M_{\pbh}$ and $\mu=\frac{m_\star M_{\pbh}}{m_\star +M_{\pbh}}$ and $b_{90}=\frac{\alpha}{\mu v_{\rm rel}^2}$, the impact parameter which results in a deflection of $90^{\circ}$.

We next describe $b_{\rm max}$ and $b_{\rm min}$. 
The maximum impact parameter is set by the requirement that the time of passage of the perturber be shorter than the orbital period of the star. This is because, for long scattering time, the adiabatic invariant dilutes the amount of energy transferred onto the star. For non-adiabaticity we require
\begin{align}
\frac{b_{\rm max}}{\sigma_{\pbh}}\ll \frac{R_{0,\star}}{\sigma_\star}
\end{align}
We see from Eqns.~\eqref{velstar} and \eqref{veldm} that $\frac{\sigma_{\pbh}}{\sigma_\star}\approx\frac{18}{4.45}$. Hence we conservatively choose $b_{\rm max}=3R_{0,\star}$ to satisfy the adiabaticity condition.  We only consider MACHOs passing within this $b_{\rm max}$ to be direct heating.  MACHOs passing outside of this will contribute to tidal heating.

In literature~\cite{binney2011galactic,Brandt:2016aco,Church:2018sro} the minimum impact parameter $b_{\rm min}$ is ignored and the denominator of the log in Eqn.~\eqref{eqn:heateqnspec} is typically set to $b_{90}$. However, we show that when the MACHOs are sparse, then $b_{\rm min}$, set by sampling, can exceed $b_{90}$, and hence determines the denominator of the log in Eqn.~\eqref{eqn:heateqnspec}.

We first require that 3 MACHOs pass through $b_{\rm max}$ i.e. $b_{\rm max}>b_{\rm samp}$.  Where we define $b_{\rm samp}$ to be the radius around any particular star through which exactly 3 MACHOs pass in the age of the UFD.
Here $b_{\rm samp}$ is given by solving
\begin{align}
   \pi b_{\rm samp}^2 \frac{f_{\pbh} \rho_{\rm DM}}{M_{\pbh}} v_{\pbh} T_{\rm UFD} =3
   \label{bsampdt}
\end{align}
Here $v_{\pbh}$ is the mean speed of the MACHOs and is given by $v_{\pbh}=\sqrt\frac{8}{\pi} \, \sigma_{\pbh}$.

For $T_{\rm UFD}=10^{10}~\textrm{year}$,
$b_{\rm max}>b_{\rm samp}$ for an initial stellar scale radius of $R_{0,\star}^i=2~\textrm{pc}$ translates to the requirement
\begin{align}
f_{\pbh}\gtrsim 3.7\times10^{-8}\frac{M_{\pbh}}{M_\odot}
\label{frate3}
\end{align}
If this condition is met, there are at least 3 $\pbh$s passing through the stellar distribution over the age of the UFD.

We conservatively use $b_{\rm min}=b_{\rm samp}$ in our calculations.  This makes sure that the heating rate we are calculating comes only from scattering events where $\pbh$s pass at least  $b_{\rm samp}$ away from the star.  We know that most stars will have such events during the age of the UFD.  By setting $b_{\rm min}=b_{\rm samp}$ we exclude heating from events where $\pbh$s pass closer to stars, because such events will only occur for a small fraction of the stars and so it provides a nonuniform heating of the stars.  Thus we conservatively exclude it.
In much of the parameter space of interest, $b_{\rm samp}$ is larger than $b_{\rm 90}$. 
\begin{align}
&b_{\rm samp}=2.6~\textrm{mpc} \sqrt{\frac{M_{\pbh}}{M_\odot}}\sqrt{\frac{1}{f_{\pbh}}} \nonumber \\
    &b_{90}=\frac{G M}{v^2}\approx 10^{-2} \textrm{mpc} \frac{M_{\pbh}}{M_\odot}
    \label{b90eqn}
\end{align}
and hence ignoring $b_{\rm min}=b_{\rm samp}$ in Eqn.~\eqref{eqn:heateqnspec} can overestimate the heating rate.

\begin{figure}
\centering
\begin{minipage}{.492\textwidth}
  \centering
\includegraphics[width=\linewidth]{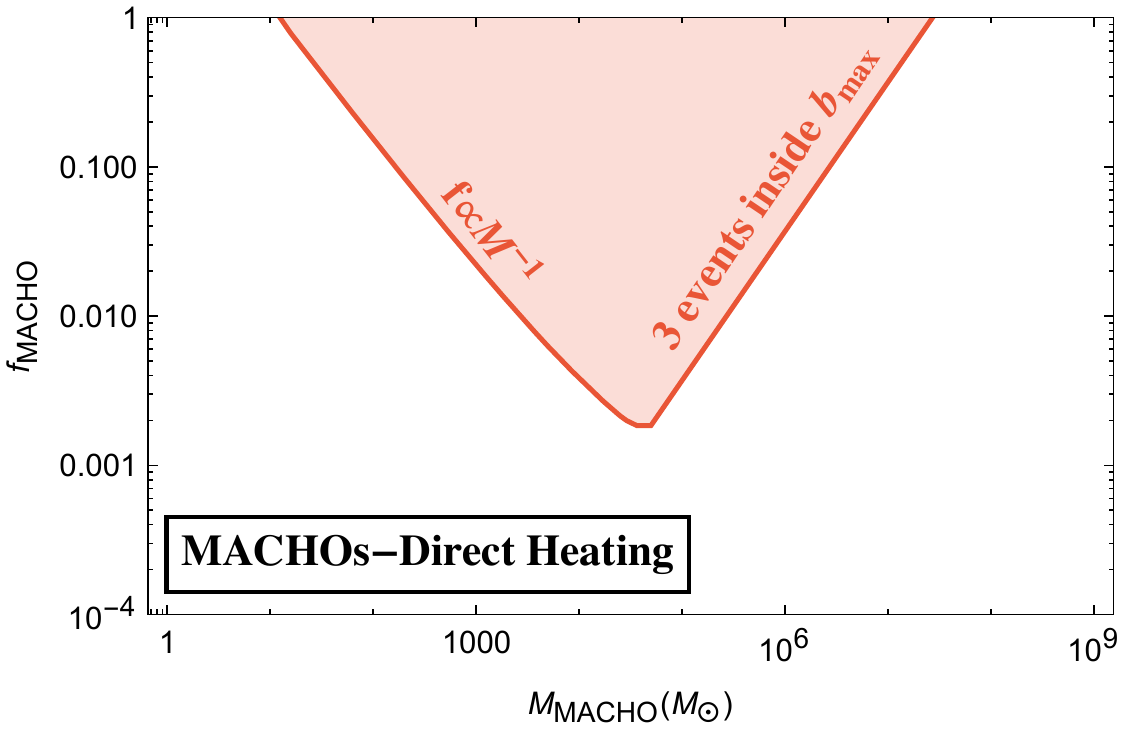}
\end{minipage}
\begin{minipage}{.492\textwidth}
  \centering
  \includegraphics[width=\linewidth]{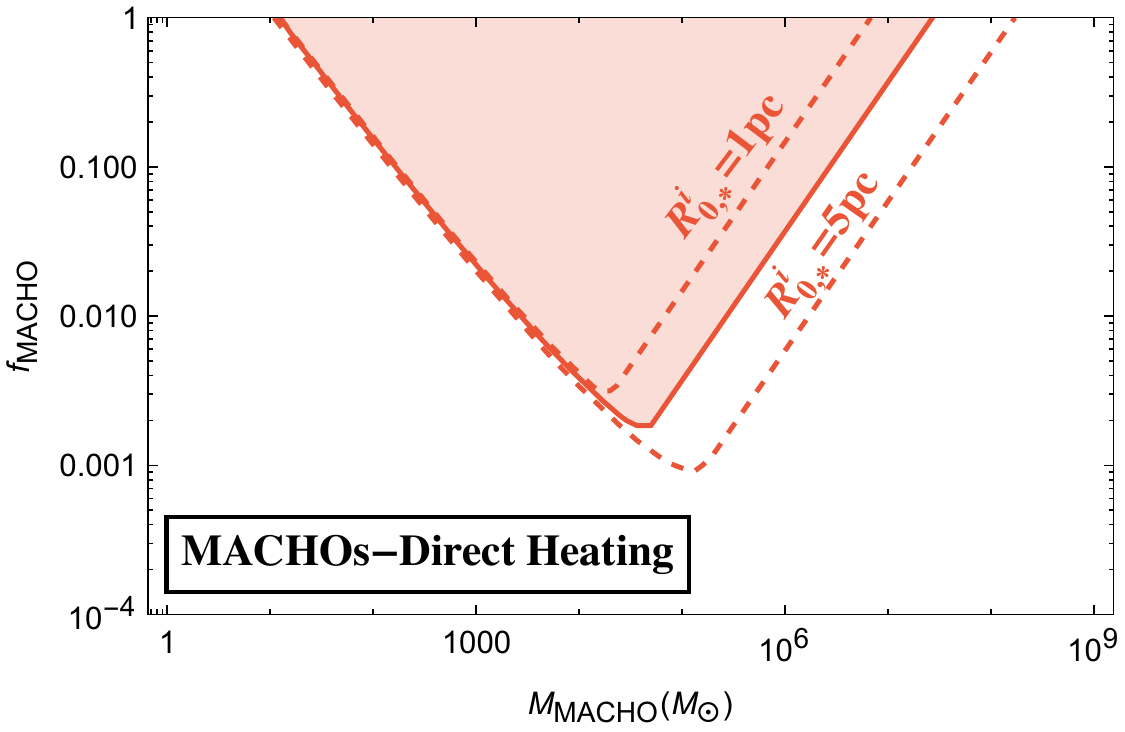}
\end{minipage}
\caption{\textbf{Left:} Constraints on $\pbh$s present in UFDs arising from the direct heating of stars showing the different regimes assuming the inital stellar half-light radius be no less than $R_{0,\star}^i=2\textrm{pc}$. \textbf{Right:} Limits corresponding to three different choices of initial stellar half-light radius $R_{0,\star}^i=1~\textrm{pc}$ and $R_{0,\star}^i=5~\textrm{pc}$ (Dashed) and $R_{0,\star}^i=2~\textrm{pc}$ (Solid shaded).}
\label{directfig}
\end{figure}

We can now solve the differential Eqn.~\eqref{heatingeqndiff}. Setting  the final stellar scale radius to be the observed $R_{0,\star}(\rm today)=24.73~\textrm{pc}$, we solve Eqn.~\eqref{heatingeqndiff} numerically to find the initial scale radius $R_{0,\star}^i\equiv R_{0,\star}(t=0)$. When this value is smaller than $2~\textrm{pc}$, we call this point excluded as shown in the shaded region in Fig.~\ref{directfig}. (The choice of $2~\textrm{pc}$ for $R_{0,\star}^i$ will be explained below). 

We describe the different regimes next. In the small MACHO mass regime, we can understand the left slope of Fig.~\ref{directfig} by obtaining an analytical solution via a series of justified approximations. When, $M_{\pbh}\ll 10^4 M_\odot$, $b_{\rm max} \gg b{\rm samp} \gg b_{90}$ and we can approximate the log in Eqn.~\eqref{eqn:heateqnspec} to be a constant, which we call $\log\Lambda^2$. Further, $\sigma_\star \ll \sigma_{\pbh}$. We will also fix $\log \left(\frac{R_{\rm UFD}}{R_{0,\star}}\right)\approx 3.5$, just to get an analytical result. As a result, the solution to Eqn.~\eqref{heatingeqndiff} simplifies to,

\begin{align}
    R_{0,\star}[t_{\rm UFD}]^2&\approx 0.34 f_{\pbh} \frac{GM_{\pbh}}{\sigma_{\rm DM}} t_{\rm UFD} \log \Lambda  + (R_{0,\star}^i)^2 \nonumber \\
   &\approx\left(24 \textrm{pc}\right)^2\frac{t_{\rm UFD}}{10^{10} {\rm year}} \frac{f_{\pbh}}{1}\frac{M_{\pbh}}{10 M_\odot}\frac{\log \Lambda}{10}\frac{18 \rm km/s}{\sigma_{\pbh}}+(R_{0,\star}^i)^2 
   \label{hlrexp}
\end{align}
where $R_{0,\star}^i$, is the initial stellar scale radius. Later in this section, we discuss the choice of $R_{0,\star}^i$, with all choices being much smaller than $R_{0,\star}=24.7~\textrm{pc}$ today. Hence, to understand the results in Fig.~\ref{directfig}, we can ignore the $(R_{0,\star}^i)^2$ term in the RHS of Eqn.~\eqref{hlrexp}.

As seen in Eqn.~\eqref{hlrexp}, starting with extremely small $R_{0,\star}^i$, the MACHO heating results in $R_{0,\star}$ exceeding the observed half-light radius when $f_{\pbh} M_{\pbh} > 10 M_{\odot}$.  
This results in the roughly $f_{\pbh} \propto M_{\pbh}^{-1}$ behavior in Fig.~\ref{directfig}. 

Next, imposing Eqn.~\eqref{frate3} for three passes of MACHOs within $b_{\rm max}$ sets the turn-around around $10^4 M_\odot$ and the $f_{\pbh} \propto M_{\pbh}$ behavior of the right side of the excluded region. \\ 

\textbf{Initial conditions}: We next turn to quantifying the uncertainty associated with the choice of the initial stellar scale radius. Eqn.~\eqref{hlrexp} naively implies an initial-condition independent, final half-light radius that scales as $t_{\rm UFD}$. While this is approximately true, the initial conditions play a big role in the determination of the $\log$ due to sampling. Afterall, an arbitrarily small $R_{0,\star}^i$ would result in no direct heating, since we set $b_{\rm max}=3R_{0,\star}$. Hence, the effect of the heating from MACHOs can be absorbed by choosing a perversely small $R_{0,\star}^i$. Numerical simulations however predict even larger half-light radii than what is observed today in low luminosity UFDs. Ref.~\cite{revaz2023compactness} compiles a list of observations and numerical simulations to exhibit the poor agreement between them at low luminosities. 
The presence of MACHO heating only exacerbates this simulation-observation discrepancy. To be conservative we choose $R_{0,\star}^i=2\textrm{pc}$ which corresponds to an initial stellar core density of $\rho_\star^i\approx5.27 M_\odot \textrm{pc}^{-3}$. This density is comparable to some of the densest stellar objects: globular clusters. In Fig.~\ref{directfig}(Right), we show the constraints on the MACHO parameter space for different choices of $R_{0,\star}^i$ in dashed contours. The central, shaded region corresponds to $R_{0,\star}^i=2~\textrm{pc}$. As explained earlier, the left edge of the plots are heating rate limited and hence are relatively agnostic to the initial conditions. The right edges are sampling limited and hence more sensitively dependent on the choice of $R_{0,\star}^i$. 
Although Fig.~\ref{directfig}(Right) shows some dependence on initial stellar scale radius, in fact in our final answer for the total excluded region from all heating will not depend on this choice as we will see below.

\subsection{Tidal Heating}
We restricted ourselves to impact parameters $b\lesssim 3 R_{0,\star}$ in the previous subsection. As the impact parameter gets larger, the direct acceleration on individual stars gets more coherent, which results in the acceleration of the star cluster as a whole. The increase in the half-light radius which is brought about by accelerating individual stars is instead caused by tidal effects. This is dealt with in this subsection.

We closely follow Ref.~\cite{vandenBosch:2017ynq} to quantify the tidal heating. Consider a MACHO that follows a path $R(t)=\left\{b,v_{\pbh}t,0\right\}$, in cartesian coordinates, with $v_{\rm rel}$, the relative velocity between the star cluster and an individual MACHO and the center of the star cluster being the origin of the coordinate system. We approximate this as a straight line because we will be interested in impact parameters $b$ (the closest approach of the orbit to the stars) that are much smaller than the average MACHO orbital radius. The deviations from this approximation were shown to be $\mathcal{O}(1)$ in Ref.~\cite{gnedin1999tidal}. 
We employ the distant tide approximation i.e. the impact parameter is larger than the size of the star system  $b\gg R_{0,\star}$. After the MACHO passes by the stars, the velocity change of an individual star located at coordinates $\{x,y,z\}$ is
\begin{align}
    \Delta v=\frac{2G M_{\pbh}}{b^2 v_{\pbh}}\left\{x,0,-z\right\}
\end{align} 
Then, the energy injected into the star cluster per unit stellar mass is,
\begin{align}
    \frac{\Delta E_{\rm dt}}{M_s}=\frac{4G^2M_{\pbh}^2}{b^4v_{\pbh}^2}\frac{\langle r^2 \rangle_\star}{3}
\end{align}
where, 
\begin{align}
    \langle r^2 \rangle_\star=\frac{1}{M_\star}\int^{r_{\rm max}}_0 4\pi r^4 \rho_\star(r)dr
    \label{rsqint}
\end{align}.

Note that this quantity diverges for large $r_{\rm max}$ owing to the fact that tidal forces grow with the radial position. We cut off the radial integral at $r_{\rm max}=2.5 R_{0,\star}$ owing to the fact that there few stars outside this radius and thus heating those distant stars should not affect the stellar scale radius much.

The tidal heating rate is
\begin{align}
H_{\star, \rm dt}=\int_{b_{\rm min}}^{b_{\rm max}} \Delta E_{\rm dt}(b) \, A_{\rm corr}(b) \, 2\pi b \, db \, \frac{\rho_{\pbh}}{M_{\pbh}} v_{\pbh} 
\label{eqn:hdt}
\end{align}
Here $A_{\rm corr}=\left(1+\left(\frac{\sigma_\star(R_{0,\star})}{R_{0,\star}} \frac{b}{\sigma_{\pbh}}\right)^2\right)^{-\frac{3}{2}}$ is the adiabatic correction factor~\cite{vandenBosch:2017ynq} and the subscript $\textrm{dt}$ denotes distant tide approximation. This exists to account for the fact that the adiabatic theorem dilutes the heating rate for scattering times that exceed the orbital frequency of the stars.  

The lower limit of the integral is given by, 
\begin{align}
    b_{\rm min}=\textrm{Max}\left(b_{\rm samp},3 R_{0,\star}\right)
\end{align}
Here, $b_{\rm samp}$ is defined in Eqn.~\eqref{bsampdt}. This ensures impact parameters are no smaller than $3 R_{0,\star}$ which (a) exceeds the $2.5 R_{0,\star}$ cutoff of the radial integral, and (b) is completely disjoint from the $b_{\rm max}=3~R_{0,\star}$ set for direct heating in Section.~\ref{secdirectheating}. For tidal heating $b_{\rm max}$ is set to $446.3~\textrm{kpc}$, the size of the UFD, but it has negligible impact on the results.

\begin{figure}
\centering
\begin{minipage}{.5\textwidth}
  \centering
\includegraphics[width=.99\linewidth]{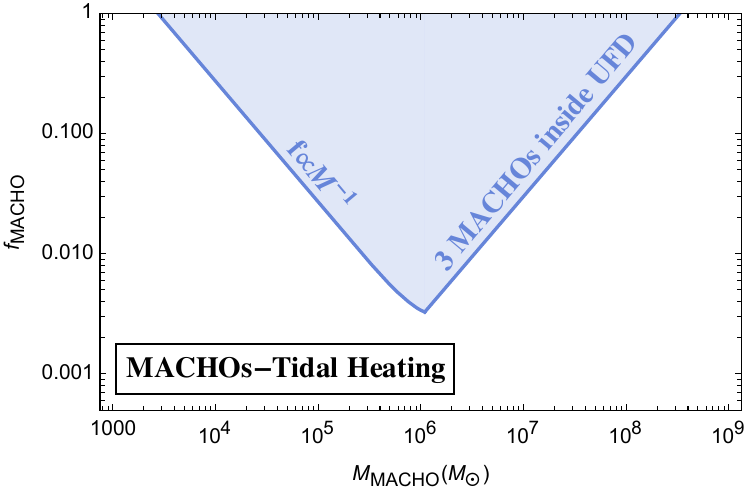}
\end{minipage}%
\begin{minipage}{.5\textwidth}
  \centering
\includegraphics[width=.99\linewidth]{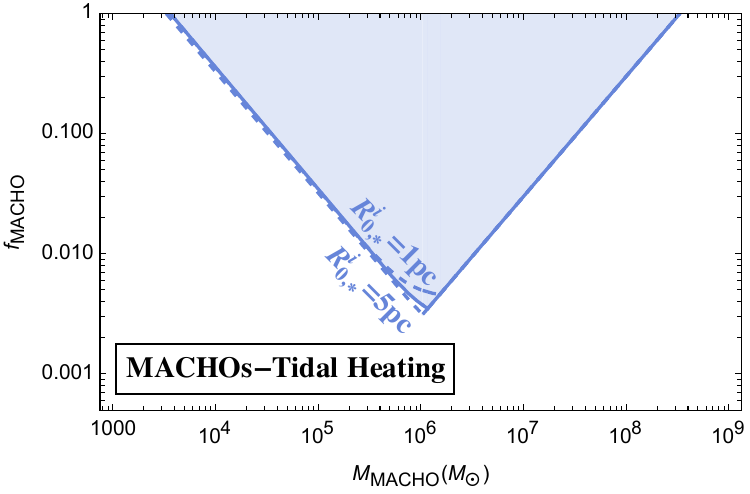}
\end{minipage}
\caption{\textbf{Left:}~Constraints on $\pbh$s present in UFDs arising from the tidal heating of stars showing the different regimes assuming the initial stellar half-light radius be no less than $R_{0,\star}^i=2\textrm{pc}$. \textbf{Right:} Limits corresponding to three different choices of initial stellar half-light radius $R_{0,\star}^i=1~\textrm{pc}$ and $R_{0,\star}^i=5~\textrm{pc}$ (Dashed) and $R_{0,\star}^i=2~\textrm{pc}$ (Solid shaded).}
\label{tidalfig}
\end{figure}

The result of solving the differential Eqn.~\eqref{heatingeqndiff} with the heating rate given by Eqn.~\eqref{eqn:hdt} results in exclusion in $f_{\pbh}$ as a function of $M_{\pbh}$ which is plotted in Fig.~\ref{tidalfig}. We next unpack the different regimes in this figure. 

For the smallest masses of interest, $b_{\rm samp}< 3 R_{0,\star}$ and hence $b_{\rm min}=3 R_{0,\star}$. For such close encounters, $A_{\rm corr}\rightarrow 1$ and we can safely ignore it in Eqn.~\eqref{eqn:hdt}. In this limit, we see that $H_{\star,\rm dt}\propto f_{\pbh} M_{\pbh}$. This results in the $f_{\pbh}\propto M_{\pbh}^{-1}$ scaling on the left hand side of Fig.~\ref{tidalfig}(Left). 

Furthermore, we see in Fig.~\ref{tidalfig}(Right) that the LHS has negligible dependence on $R_{0,\star}^i$. We can see this from the functional form of $H_{\star,\rm dt}$ in the $A_{\rm corr}\rightarrow 1$ and $b_{\rm min}\ll b_{\rm max}$ limit,   
\begin{align}
   H_{\star,\rm dt} &\propto \int_{b_{\rm min}}^{b_{\rm max}}\frac{1}{b^4} \langle r^2 \rangle_\star \, 2 \pi b \, db  \nonumber \\
   &\propto \frac{\langle r^2 \rangle_\star }{b_{\rm min}^2}
\end{align}
Furthermore, since $\langle r^2 \rangle_\star \propto (R_{0,\star})^2$ and $b_{\rm min}=3 R_{0,\star}$ in this regime, the heating rate is independent of the choice of $R_{0,\star}^i$. Hence, the LHS independence of $R_{0,\star}^i$ in Fig.~\ref{tidalfig}(Right).

For large enough masses, $\approx10^6 M_{\odot}$, $b_{\rm samp}$ crosses $3 R_{0,\star}$, and $b_{\rm min}$ is set by $b_{\rm samp}$. This results in the turn around. 

Finally, the right edge in Fig.~\ref{tidalfig}(Left) is set by the requirement that there are 3 or more MACHOs inside the UFD, i.e. $f_{\pbh} M_{\rm UFD}>3M_{\pbh}$. This condition is $R_{0,\star}^i$ independent as well, and hence in Fig.~\ref{tidalfig}(Right), the $R_{0,\star}^i$ dependence only shows up at the bottom tip of the contours.   

\subsection{Combined results}
\begin{figure}
\centering
\begin{minipage}{.5\textwidth}
  \centering
\includegraphics[width=.99\linewidth]{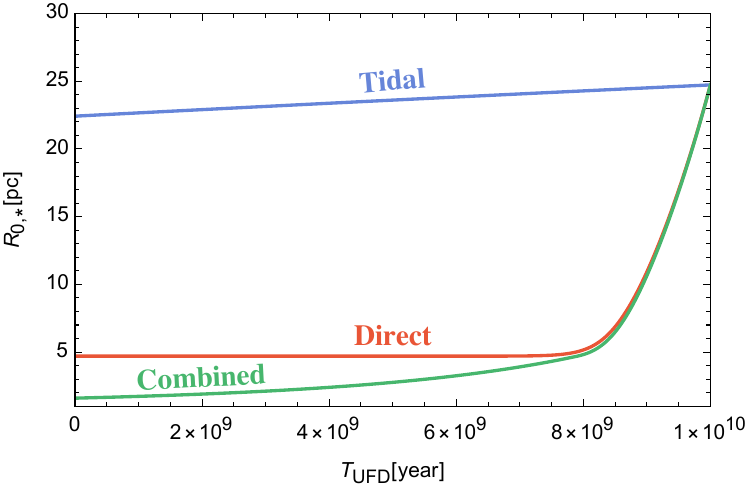}
\end{minipage}%
\begin{minipage}{.5\textwidth}
  \centering
\includegraphics[width=.99\linewidth]{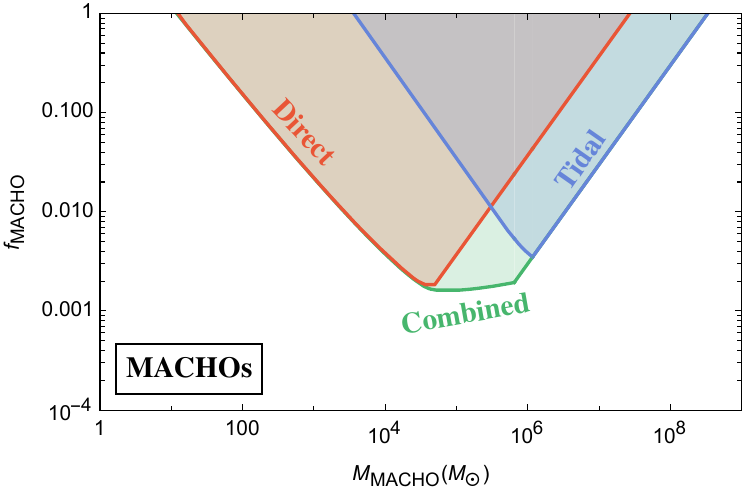}
\end{minipage}
\caption{\textbf{Left:} The evolution of the stellar scale radius $R_{0,\star}$ is shown vs the age of the UFD, $T_{\rm UFD}$. Evolution from tidal (direct) heating alone is shown in blue (red), and the combined heating in green for a sample point $M_{\pbh}=3\times10^5$ and $f_{\pbh}=2\times10^{-3}$. \textbf{Right:} Limits from direct (tidal) heating on the fraction of DM in MACHOs is shown as a function of the MACHO mass for direct (tidal) heating alone in red (blue) and combined heating (green) on $\pbh$s assuming the initial stellar half-light radius be no less than $R_{0,\star}^i=2\textrm{pc}$.}
\label{fig:combinedleftright}
\end{figure}

While we discussed the limits from each regime individually, we can add the direct and tidal heating contributions since they correspond to non-overlapping impact parameter regions. Hence, one can solve Eqn.~\eqref{heatingeqndiff} with $H_\star=H_{\star,\rm direct}+H_{\star,\rm dt}$, with  $H_{\star,\rm direct}$ given in Eqn.~\eqref{eqn:heateqnspec} and  $H_{\star,\rm dt}$ given in Eqn.~\eqref{eqn:hdt}. The result of this procedure is displayed in Fig.~\ref{fig:combinedleftright}. 

In Fig.~\ref{fig:combinedleftright} (left) the time dependent growth of the stellar scale radius $R_{0,\star}$ is shown as a function of the age of the UFD for direct heating (Red), tidal heating (Blue) and the effect of adding up both heating rates in Green for a sample point $M_{\pbh}=3\times10^5$ and $f_{\pbh}=2\times10^{-3}$. For all three curves the scale radius today is fixed to $R_{0,\star}\left(T_{\rm UFD}=10^{10}~\textrm{year}\right)=24.73~\textrm{pc}$.  Then the heating equation is solved backwards in time to predict the initial stellar scale radius $R_{0,\star}^i$. One can see that tidal heating has the smallest effect on the scale radius, while direct heating alone only predicts $R_{0,\star}^i\approx 5~\textrm{pc}$ . This is because for $R_{0,\star}< 5~\textrm{pc}$, direct heating is sampling limited for this particular choice of $\{M_{\pbh},f_{\pbh}\}$. Tidal heating is too weak by itself as well. However, even for $R_{0,\star}^i\approx 1~\textrm{pc}$, tidal heating can cause limited heating up to $5~\textrm{pc}$ where direct heating takes over to expand to $24.73~\textrm{pc}$. 
The excluded region from combined heating
is plotted in Fig.~\ref{fig:combinedleftright}. (right) in green. It excludes new parameter space not excluded by direct heating or tidal heating alone.

\section{Migration of MACHOs in the UFD}
\label{sec:migration}
\begin{figure}
\centering
\begin{minipage}{.49\textwidth}
  \centering
\includegraphics[width=.99\linewidth]{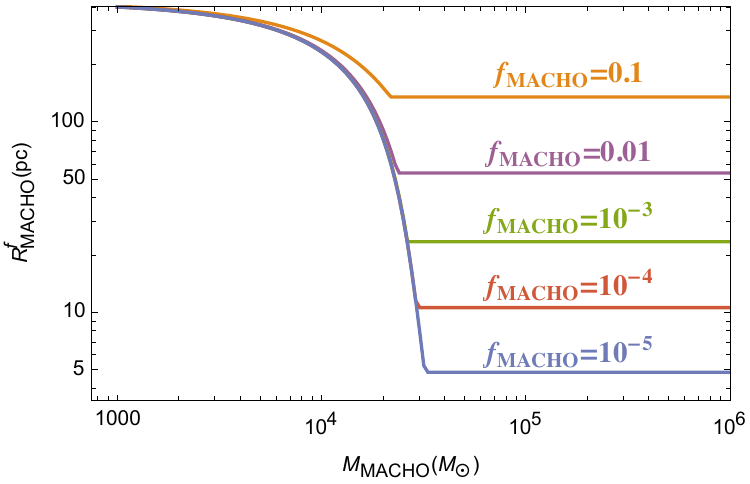}
\end{minipage}
\begin{minipage}{.49\textwidth}
  \centering
\includegraphics[width=.99\linewidth]{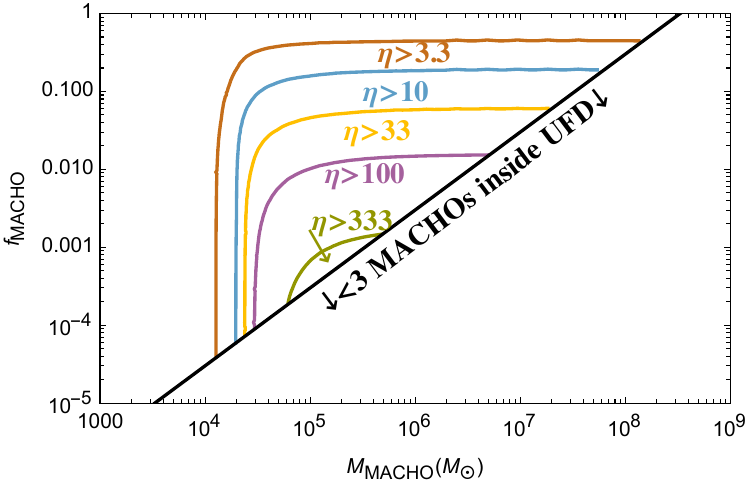}
\end{minipage}
\caption{\textbf{Left:}  The MACHO scale radius at time $t=10^{10} \textrm{year}$, $R_{\pbh}^f$, is shown as a function of MACHO mass for different values of MACHO DM fraction. \textbf{Right:} Contours of constant resultant enhancement $\eta$ (defined in text) are shown in the MACHO fraction vs MACHO mass plane.}
\label{fig:migrationenh}
\end{figure}
Until now we have assumed that the $\pbh$ profile is the same as the smooth DM profile for our heating calculations. However, we have ignored the heat exchange between MACHOs and the smooth DM. In this section, we consider the cooling of MACHOs due to their heat transfer to the smooth DM, which results in the contraction of the MACHO scale radius $R_{\pbh}$. 
We will assume that the MACHOs continue to maintain a Dehnen profile as they contract i.e.
\begin{align}
    \rho_{\pbh}(r)=f_{\pbh}\frac{\left({3-\gamma}\right) M_{\rm UFD}}{4\pi R_{\pbh}^3} \left(\frac{r}{R_{\pbh}}\right)^{-\gamma} \left(1+\frac{r}{R_{\pbh}}\right)^{\gamma-4}
    \label{eqn:rhomacho}
\end{align}
with $R_{\pbh}$ now being time-dependent and we take $\gamma=0$.

The $\pbh$ velocity dispersion can be estimated from requiring hydrostatic equilibrium.  We use the exact answer in our calculations.  However in order to aid in understanding, here we will just give the expression in the $R_{\pbh}\ll R_{\rm UFD}$ and $r\ll R_{\rm UFD}$ limit:
\begin{align}
    \sigma^2_{\pbh}(r)&\approx \frac{(1-f_{\pbh}) M_{\rm UFD}(r+R_{\pbh}) (3 r+R_{\pbh})}{6R_{\rm UFD}^3}+\frac{f_{\pbh}M_{\rm UFD}  (6 r+R_{\pbh})}{30 (r+R_{\pbh})^2} \nonumber \\
    &\approx \frac{(1-f_{\pbh}) M_{\rm UFD}R_{\pbh}^2}{6R_{\rm UFD}^3}
    \label{eqn:veldismigration}
\end{align}
where in the second line we have taken the limit $r \ll R_{\pbh}$
and $f_{\pbh}\ll1$.
We can now follow the prescription developed in Section.~\ref{sec:exphlr} to calculate the time dependence of $R_{\pbh}$. 

Similar to Eqn.~\eqref{heatint}, we start by equating the time derivative of the total MACHO energy $\mathcal{E}_{\pbh}$ given in Appendix.~\ref{totenergy} to the total heating rate, i.e.
\begin{align}
     \frac{d\mathcal{E}_{\pbh}(R_{\pbh})}{dt}= \int dr \, 4\pi r^2  \rho_{\pbh}(r) \, H_{\pbh} \left( \sigma_{\pbh}(r),\sigma_{\rm DM}(r),\rho_{\rm DM}(r) \right) 
    \label{heatintmacho}
\end{align}
where $H_{\pbh}$ is the heating rate of MACHOs per unit mass and is given by,

\begin{align}
    H_{\pbh}
    =2\sqrt{2\pi} G^2  \left(1- f_{\pbh}\right)\rho_{\rm DM} \frac{ \left(m_{\rm DM} \sigma_{\rm DM}^2-M_{\pbh}\sigma_{\pbh}^2\right)}{\left(\sigma_{\rm DM}^2+\sigma_{\pbh}^2\right)^\frac{3}{2}}&\log \left(\frac{\alpha^2+b_{\rm max}^2 \mu^2
   v_{\rm rel}^4}{\alpha^2+b_{\rm min}^2 \mu^2
   v_{\rm rel}^4}
   \right)\nonumber \\
 = ... &\log \left(\frac{b_{90}^2+b_{\rm max}^2}{b_{90}^2+b_{\rm min}^2 }\right)
    \label{heateqndirmacho}
\end{align}
where $m_{\rm DM}$ is the individual particle mass of the smooth DM component. Since $M_{\pbh}\gg m_{\rm DM}$, $H_{\pbh}\propto M_{\pbh}$ and hence the effect of migration increases with $M_{\pbh}$. 

There are no sampling issues here since the DM is considered to be smooth. Hence we can safely set $b_{\rm min}\rightarrow 0$. 
We assume that $R_{\pbh}=R_{\rm UFD}$ at $t=0$ and solve the differential Eqn.~\eqref{heatintmacho} to obtain $R_{\pbh}(t)$. Note that $R_{\rm UFD}$ is held fixed through this exercise since we are ignoring the back-reaction onto the smooth DM component. This assumption breaks down when the MACHO density becomes comparable to the smooth DM density.  In this whole section we consider $f_{\pbh} \ll 1$, so the assumption is initially valid. But as the MACHO distribution contracts, the local MACHO density could become comparable to the smooth DM density and the heat transfer would then have a significant effect on the smooth DM. To address this, we stop the $R_{\pbh}(t)$ contraction when the total $\pbh$ mass enclosed inside $R_{\pbh}(t)$ equals the smooth DM mass enclosed inside the same radius. Setting $R_{\pbh}^f=R_{\pbh}(t=10^{10}~\textrm{year})$, the above condition translates to $R_{\pbh}^f$ being cutoff at $R_{\pbh}^{\rm cutoff}$ given by,
\begin{align}
    R_{\pbh}^{\rm cutoff}\approx R_{\rm UFD} f_{\pbh}^\frac{1}{3}
    \label{rcut}
\end{align}
The result of solving Eqn.~\eqref{heatintmacho} for $R_{\pbh}(t)$ is shown in Fig.~\ref{fig:migrationenh}~(Left). In Fig.~\ref{fig:migrationenh} (Left) we plot  $R_{\pbh}^f=R_{\pbh}(t=10^{10}~\textrm{year})
$ as a function of $M_{\pbh}$. As seen in the figure, the effect of migration is negligible for masses much below $10^4 M_{\odot}$ and it increases with increasing $M_{\pbh}$. The effect of cutting off migration when MACHO density matches the ambient DM density translates to the eventual flat asymptote i.e. $R_{\pbh}^f=R_{\pbh}^{\rm cutoff}$ whose $f_{\pbh}$ dependence is given in Eqn.~\eqref{rcut}. 

We see that migration can play a significant effect in much of the parameter space especially when $M_{\pbh}\gtrsim 10^{4} M_\odot$. This can in-turn result in higher heat transfer to stars as compared to the estimates made in Section.~\ref{sec:exphlr}. The precise way to take this into account is by replacing the time-independent $\rho_{\pbh}$ in Eqn.~\eqref{eqn:heateqnspec} (Direct Heating) and Eqn.~\eqref{eqn:hdt} (Tidal Heating), with a time-dependent $\rho_{\pbh}(t)$ whose time-dependence is derived from the time dependence of $R_{\pbh}$ as given in Eqn.~\eqref{eqn:rhomacho}. However, this is computationally intensive and we defer this exercise to future work. 
Instead, we approximate this effect by replacing $f_{\pbh}$ used to calculate the heating rate in previous sections by $f_{\pbh} \times \eta$, where 
\begin{align}
\eta(R)=\frac{1}{T_{\rm UFD}}\int_0^{T_{\rm UFD}} dt \frac{\rho_{\pbh}(R)}{\rho_{\rm DM}(R)}
\end{align}
We are interested in $\eta$ evaluated near the stellar scale radius, but we find that $\eta$ is only weakly dependent on the choice of $R$ when evaluated in the range $2~\textrm{pc}<R<24.7~\textrm{pc}$ for the parameters of interest. While this prescription involving the time-averaged enhancement $\eta$ might appear crude, we have checked its accuracy by comparing it to the precise calculation in a handful of parameter points to find that this is an excellent approximation.

\begin{figure}
\centering
\begin{minipage}{.5\textwidth}
  \centering
\includegraphics[width=.99\linewidth]{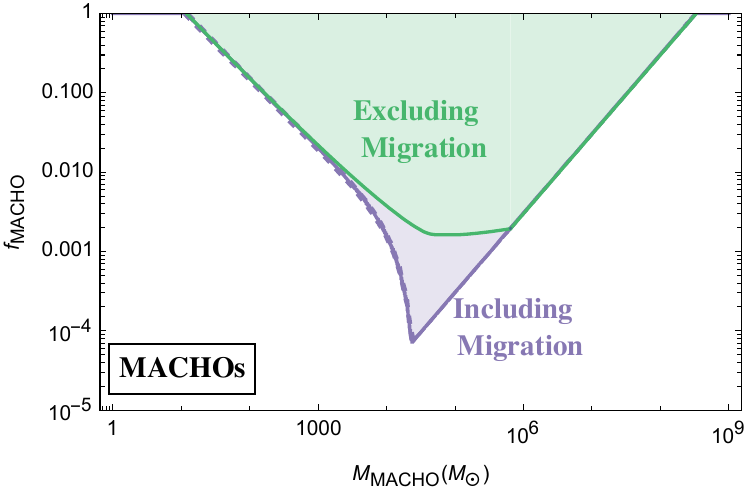}
\end{minipage}%
\begin{minipage}{.5\textwidth}
  \centering
\includegraphics[width=.99\linewidth]{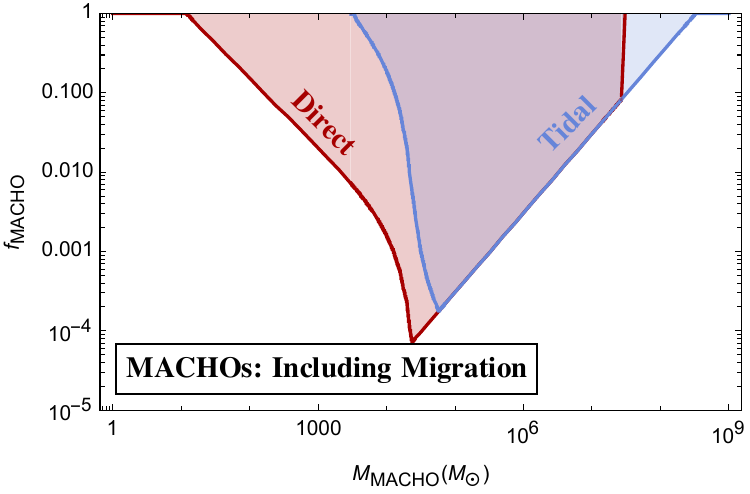}
\end{minipage}
\caption{\textbf{Left:} Limits on the MACHO fraction as a function of MACHO mass are shown including the effect of migration of MACHOs due to heat transfer with the smooth DM in purple, and excluding migration in green (the green curve is identical to the one in Fig.~\ref{fig:combinedleftright}(right)). While the solid purple curve corresponds to $R_{0,\star}^i=2~\textrm{pc}$, also shown in dashed purple are  $R_{0,\star}^i=1~\textrm{pc}$ and $R_{0,\star}^i=5~\textrm{pc}$. There is no discernable difference between the three choices. 
\textbf{Right:} Limits derived solely from direct heating (Red) and tidal heating (Blue).}
\label{fig:migcomp}
\end{figure}

Solving the differential equations in Section.~\ref{sec:exphlr} after including the enhancement due to migration, we find improved exclusions, which are plotted in purple in Fig.~\ref{fig:migcomp}. In Fig.~\ref{fig:migcomp} (Left), the steep increase in density enhancements above $M_{\pbh}=10^4 M_{\odot}$ seen in Fig.~\ref{fig:migrationenh}, translate to steep improvement in limits on $f_{\pbh}$ in this mass range. This produces the purple region over and above the green region from Section~\ref{sec:exphlr}. However the improvement in limits is cutoff by our requirement of three MACHOs inside the UFD which  leads to the RHS of the plot. While the initial stellar scale radius is chosen to be $R_{0,\star}^i=2~\textrm{pc}$, we also show contours of $R_{0,\star}^i=1~\textrm{pc}$ and $R_{0,\star}^i=5~\textrm{pc}$ in purple dashed lines. There is no discernable difference between these choices. Hence our final exclusion results are relatively agnostic to the choice of $R_{0,\star}^i$. 

The reason for this can be understood from analyzing Fig.~\ref{fig:migcomp} (Right), where we break-down the exclusions from direct (Red) and Tidal (Blue) heating. We see that exclusions from direct heating is significantly expanded after taking into account migration when compared to limits in Fig.~\ref{fig:combinedleftright} which did not include the effect of migration. As argued earlier, the only dependence on $R_{0,\star}^i$ arose in the right edge of the direct heating region where the events were sampling limited. After including migration, the right edge of the direct heating region is either deep into the ``3 MACHO inside UFD" region or in the region where tidal heating is efficient. Hence our final result are relatively independent of the initial scale radius $R_{0,\star}^i$.

\begin{figure}[h]
     \centering
    \includegraphics[scale=0.99]{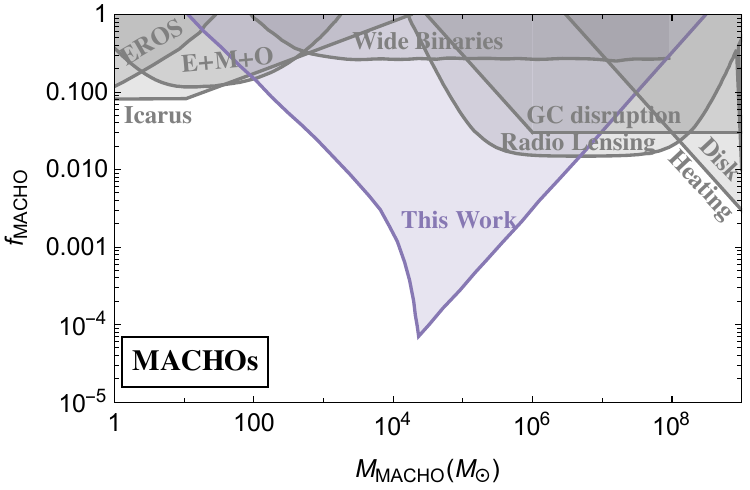}
   
     \caption{Limits on the MACHO fraction as a function of MACHO mass are shown in purple. Other limits on this region come from microlensing (EROS)~\cite{EROS-2:2006ryy} and combined EROS/MACHO/OGLE~{Blaineau:2022nhy}, caustic crossing of Icarus~\cite{Oguri:2017ock} and multiple imaging of radio sources~\cite{Wilkinson:2001vv, 10.1093/mnras/stac915}. Also shown are limits derived from the non-disruption of wide binaries~\cite{Ramirez:2022mys}, from the non-disruption of galaxy clusters~\cite{moore1993upper} and the non-observation of excess disk heating~\cite{Carr:2020gox}. }
      \label{moneyplot} 
\end{figure}

In Fig.~\ref{moneyplot}, we compare this limit with other existing limits on MACHOs which are shown in gray. The existing limits derived from microlensing by the EROS collaboration~\cite{EROS-2:2006ryy} and combined EROS/MACHO/OGLE~{Blaineau:2022nhy}, caustic crossings with  Icarus~\cite{Oguri:2017ock} and multiple imaging of compact  radio sources~\cite{Wilkinson:2001vv, 10.1093/mnras/stac915} are shown. Dynamical heating limits derived from the non-disruption of wide binaries~\cite{Ramirez:2022mys}, globular clusters~\cite{moore1993upper} and disk heating~\cite{Carr:2020gox} are also shown. It is important to point out that these do not include the effect of migration and the limits could get stronger upon incorporation of the enhancement from migration in galaxies.
Migration was also considered in the context of galactic halos in~\cite{carr1999dynamical} where bounds were set based on MACHO migration, causing mass in the galactic nucleus to exceed what is observed. Their estimation of migration suffers from two problems and hence we do not show this bound. First, radial orbits are assumed for the initial MACHO population even well inside the MACHO virial radius. We find that this overestimates the effect of migration compared to assuming a hydrostatic equilibrium and an isotropic population instead. Second, and more importantly, Ref.~\cite{carr1999dynamical} acknowledges that taking into account the back-reaction onto the smooth DM population could arrest over-densities in MACHOs from ever exceeding the local smooth density. We have taken this effect into account by cutting off migration as discussed in and around Eqn.~\eqref{rcut}. Finally, we do not show limits that are specific to PBHs and might not apply to generic MACHOs, such as ones derived from gravitational waves~\cite{Kavanagh:2018ggo,LIGOScientific:2019kan} and accretion~\cite{accretion1,accretion2,accretion3,accretion4}. Although Ref.~\cite{Bai:2020jfm} works out accretion for extended objects, they work in the $f_{\pbh}=1$ regime and hence an $f_{\pbh}$ vs $M_{\pbh}$ exclusion is not easily derivable from their work.

As seen, the limits set from this work improve existing limits on MACHOs in the $100 M_{\odot}$ to $10^7 M_{\odot}$ mass range. Just above $M_{\pbh}=10^4 M_{\odot}$, MACHO fractions as small as $f_{\pbh}=10^{-4}$ are ruled out. 

\section{Conclusion}
\label{sec:conclusion}
MACHOs are a promising dark matter candidate and a consequence of a plethora of BSM theories. Their cosmic presence or lack thereof can tell us about the validity of the underlying BSM theories as well as the cosmology of the early universe.  In this paper we use ultrafaint dwarf galaxies to set limits on MACHOs (and PBHs). These constraints arise due to heat transfer between the gas of MACHOs and the gas of stars, resulting in an expansion of the stellar scale radius that exceeds observations. 

We point out important corrections to the heat transfer rate equation used in previous work. And we extend the reach to MACHO masses higher than $10^4 M_\odot$, calculating the cutoff of the effect at higher masses. To our knowledge, we present the first analysis of heat transfer between the MACHOs and the smooth dark matter of ultrafaint dwarfs resulting in MACHO migration towards the dwarf's center. We also incorporated this into the heating of stars which improves the limit around the $10^4~M_\odot$ range.
This work sets the most stringent limits derived to date on the dark matter fraction in MACHOs over several decades in MACHO mass. Other dynamical constraints such as the heating of wide binaries~\cite{Ramirez:2022mys}, globular clusters~\cite{moore1993upper} and disk heating~\cite{Carr:2020gox} might also become stronger after taking into account a similar migration effect for MACHOs in galaxies. We leave this for future work. 

Dynamical heating is also relatively agnostic towards the spatial extent of these MACHOs. In a companion paper~\cite{pwgforthcoming}, we analyze the degrading of these constraints for intra-galactic substructure that is more extended and diffuse than MACHOs and set stringent limits on the primordial power spectrum that would result in such substructure.

In this work, we took an analytical approach to this problem and hence made several conservative assumptions along the way. We hope that more robust simulations in the style of ~\cite{Stegmann:2019wyz,zhu2018primordial}, but which additionally incorporate the effect of migration of MACHOs can relax these assumptions and improve on the limits set in this paper. 

With advancements in telescope capabilities as well as developments in our theoretical understanding of galaxy formation, improvements can be made to our analysis to set better limits or possibly even discover MACHOs in ultrafaint dwarfs using the heat transfer effect described in this paper. For example our assumption about the total dwarf halo mass can be made robust via observations of stars in the periphery \cite{chiti2021extended} of the ultrafaint dwarf. This could also improve limits or possibly allow us to observe the effect of tidal heating which  disproportionately heats the outer layers of the star cluster which are the most loosely bound. Another discovery direction is the possible future observation of correlations between the stellar scale radii and the dark matter density and velocity dispersion of  ultra-faint dwarf galaxies, which would hint at the possibility of  dynamical heating from $\pbh$s setting the observed stellar scale radii. 

\acknowledgments

We thank Z.~Bogorad for useful discussions.

The authors acknowledge support by NSF Grants PHY-2310429 and PHY-2014215, Simons Investigator Award No.~824870, DOE HEP QuantISED award \#100495, the Gordon and Betty Moore Foundation Grant GBMF7946, and the U.S. Department of Energy (DOE), Office of Science, National Quantum Information Science Research Centers, Superconducting Quantum Materials and Systems Center (SQMS) under contract No.~DEAC02-07CH11359.

\appendix
\section{Total energy attributed to an individual population}
In this paper, we equated the heating rate transferred onto population A to the rate of change of total energy of population A, $\mathcal{E}_A$ i.e $H_A=\frac{1}{M_A} \frac{d\mathcal{E}_A}{dt}$.  In this section, we discuss the difficulty in unambiguously  identifying this quantity $\mathcal{E}_A$ the ``total energy of population A".

We start the discussion by first deriving equations for the kinetic and potential energies. The velocity dispersion for species $A$ in hydrostatic equilibrium is given by
\label{totenergy}
\begin{align}
\sigma_A^2(R)=\frac{G}{\rho_A(R)} \int_R^\infty \rho_A(r) \frac{M_{\rm enc} (r)}
{r^2} dr
\end{align}
where $M_{\rm enc}(r)$ is the total mass enclosed (including all species) within radius r, and $\rho_A(r)$ is the mass density of species $A$. 
Then, the kinetic energy of species $A$ is given by,
\begin{align}
\textrm{KE}_A&=\frac{3}{2} T\nonumber \\&=\frac{3}{2}\int_0^\infty 4\pi R^2 \rho_A(R) \sigma_A^2(R) \nonumber \\&=\frac{3}{2}\int_0^\infty 4\pi R^2 dR \int_R^\infty G\rho_A(r) \frac{M_{\rm enc} (r)}
{r^2} dr
\end{align}

Now,
\begin{align}
\textrm{KE}_A&=\frac{3}{2}\int_0^\infty 4\pi R^2 dR \int_R^\infty G\rho_A(r) \frac{M_{\rm enc} (r)}
{r^2} dr \nonumber \\
&=\int_0^\infty 2\pi dR^3 F[R]
\end{align}
Here $F[R]=\int_R^\infty \rho(r) \frac{GM_{\rm enc}(r)}{r^2} dr$.
We can integrate by parts to obtain, 
\begin{align}
\textrm{KE}_A=(2\pi R^3F[R])^\infty_0+\int_0^\infty 2\pi R^3 \rho_A(R) \frac{GM_{\rm enc}(R)}{R^2} dR
\end{align}
Here the $+$ sign on the second term arises because $\frac{dF[R]}{dR}=-\rho[R]\frac{GM_{\rm enc}(R)}{R^2}$. 

For $R>>R_0$, the typical size of the object, let, $\rho \propto r^{-n}$. Then,
\begin{align}
  F[R\gg R_0]\propto \int_R^\infty \rho(r) \frac{M}{r^2} \propto R^{-(n+1)}
\end{align} 
Thus if $n>2$, $R^3 F[R]_{R\rightarrow \infty} =0$.

In the small $R$ limit, let $\rho \propto R^{-k}$, then,
\begin{align}
   F[R\ll R_0]\propto \int_R^\infty \rho(r) \frac{M_{\rm enc}(r)}{r^2} \propto R^{2(1-k)}
\end{align} 
Thus if $k<5/2$, $R^3F[R]_{R\rightarrow 0} =0$. As a result,

\begin{align}
\textrm{KE}_A=\int_0^\infty 2\pi R \rho_A(R) GM_{\rm enc}(R) dR
\end{align}

Hence \begin{align}
\textrm{KE}_{\rm tot}=\int_0^\infty 2\pi R \sum_i \rho_{\rm i}(R) GM_{\rm enc}(R) dR
\end{align}
where the sum over $i$ is the sum over different species. 

Note that while kinetic energies can be calculated for each species, individual potential energies are ill-defined. The total potential energy is given by

\begin{align}
    \textrm{PE}_{\rm tot}=-\int_0^\infty \frac{ G M_{\rm enc}(r)}{r} \rho_{\rm tot}(r) 4\pi r^2 dr
    \end{align}
where $\rho_{\rm tot}=\sum_i \rho_i$
    Hence,
    \begin{align}
    \textrm{PE}_{\rm tot}=-2\textrm{KE}_{\rm tot}
    \end{align}
This is the virial theorem for multiple species, it is only valid after the sum over species. Herein lies the problem, it is wrong to assign $\mathcal{E}_A=-\textrm{KE}_A$, since the virial theorem does not apply to individual species. 

In our problem we have three distinct populations, stars, $\pbh$ and the smooth DM. 
$\textrm{KE}_{\rm tot}$ has contributions from 
\begin{align}
\textrm{KE}_{\rm tot}=\textrm{KE}_\star+\textrm{KE}_\textrm{\pbh}+\textrm{KE}_\textrm{smooth}
\end{align}
Since $\rho$ and $M_{\rm enc}$ could each have contributions from these three populations, there are terms proportional to the product of two total masses, i.e. $\left\{M_{\star},M_{\pbh}^{\rm tot},M_{\rm smooth}\right\}\times \left\{M_{\star},M_{\pbh}^{\rm tot},M_{\rm smooth}\right\}$. 

\begin{align}
\textrm{KE}_\star&=GM_\star^2 F_1(R_{0,\star})+GM_\star M_{\pbh}^{\rm tot} F_2\left(R_{0,\star},R_{\pbh}\right)+GM_\star M_{\rm smooth} F_2\left(R_{0,\star},R_{\rm smooth}\right) \nonumber  \\
\textrm{KE}_{\pbh}&=G\left(M_{\pbh}^{\rm tot}\right)^2 F_3\left(R_{\pbh}\right)+GM_{\pbh}^{\rm tot}M_{\rm smooth} F_5\left(R_{\pbh},R_{\rm smooth}\right)+GM_\star M_{\pbh}^{\rm tot}F_4\left(R_{0,\star},R_{\pbh}\right)\nonumber\\
\textrm{KE}_{\rm smooth}&=GM_{\rm smooth}^2 F_3\left(R_{\rm smooth}\right)+GM_{\pbh}^{\rm tot}M_{\rm smooth} F_5\left(R_{\rm smooth},R_{\pbh}\right)+GM_\star M_{\rm smooth} F_4\left(R_{0,\star},R_{\rm smooth}\right) 
\end{align}
Here, $M_{\star}$, is the total mass in stars, $M_{\pbh}^{\rm tot}=f_{\pbh} M_{\rm UFD}$ is the total mass in $\pbh$s and $M_{\rm smooth}=\left(1-f_{\pbh}\right) M_{\rm UFD}$, is the total mass in smooth DM and,
\begin{align}
    F_1(R_{0,\star})= \frac{2\pi}{M_\star^2} \int_0^\infty dr\rho_\star(r,R_{0,\star}) \int_0^r dr' 4\pi r'^2 \rho_\star(r',R_{0,\star})
\end{align}
\begin{align}
    F_2(R_{0,\star},R)=\frac{2\pi}{M_{\pbh}^{\rm tot} M_{\rm UFD}}\int_0^\infty dr  \rho_\star(r,R_{0,\star}) \int_0^r dr' 4\pi r'^2 \rho_{\rm DM}(r',R)
\end{align}
\begin{align}
    F_3(R)=F_5(R,R')
    \end{align}
\begin{align}
    F_4(R_{0,\star},R)= \frac{2\pi}{M_\star M_{\rm smooth} }\int_0^\infty dr \rho_{\rm DM}(r,R) \int_0^r dr' 4\pi r'^2 \rho_\star(r',R_{0,\star})
\end{align}
\begin{align}
    F_5(R,R')=\frac{2\pi}{M_{\rm smooth}^2} \int_0^\infty dr \rho_{\rm DM}(r,R) \int_0^r dr' 4\pi r'^2 \rho_{\rm DM}(r',R')
\end{align}
The potential energy can be written pairwise as,
\begin{align}
\textrm{PE}_{\rm tot}=\textrm{PE}_{\star,\rm smooth}+\textrm{PE}_\textrm{\rm smooth,\pbh}+\textrm{PE}_{\pbh,\star}+\textrm{PE}_{\star,\star}+\textrm{PE}_\textrm{\pbh,\pbh}+\textrm{PE}_{\rm smooth,smooth}
\end{align}

These distributions are intially virialized, hence $2\textrm{KE}_{\rm tot}=-\textrm{PE}_{\rm tot}$. 

Let us consider the heat transfer $H_\star$ from $\pbh$s to stars first. 

We assume that this heat transfer results in increase of $R_{0,\star}$ for the stars and decrease of $R_{\pbh}$ for the $\pbh$s while $R_{\rm UFD}$ remains unaffected. 
The evolution of $R_{0,\star}$ would be straightforward if the stellar system was heated by a source that is not present in the same UFD. However, the $\pbh$s are very much part of the UFD, and the change in  $R_{0,\star}$ affects all three kinetic energies and all combinations of potential energies. 

In the case where $M_{\pbh} \ll M_{\rm smooth}$ i.e. $f_{\pbh} \ll 1$, the problem simplifies. This is because increasing $R_{0,\star}$, affects terms proportional to $M_{\rm smooth}$ more than terms proportional to $M_{\pbh}$. 

In other words, the largest effect of expanding $R_{0,\star}$ is to the potential energy of the star-smooth system, and to the kinetic energy of the stars and smooth components. The $R_{0,\star}$ dependence on the $\pbh$ terms is negligible in comparison.
In this limit, we can equate the total heating rate of stars, to all the terms in the total energy which have at-least one power of $M_\star$, i.e.

\begin{align}
    H=\frac{1}{M_\star}\frac{d\mathcal{E}_\star}{dt}
\end{align}    
    where,
    \begin{align}
    \mathcal{E}_\star=-GM_\star^2 F_1(R_{0,\star})-GM_\star M_{\rm smooth}\left(F_2\left(R_{0,\star},r_{\rm smooth}\right)+F_4\left(R_{0,\star},r_{\rm smooth}\right)\right)
\end{align}
When $f_{\pbh}\ll 1$, $M_{\rm smooth}\approx M_{\rm UFD}$. This gives,
\begin{align}
\frac{d\mathcal{E}_\star}{dt}=
    G M_\star d\left(\frac{3\pi M_\star}{64R_{0,\star}}+\frac{3 R_{0,\star}^2}{8R_{\rm UFD}^3} M_{\rm UFD} \left(2. \log\frac{R_{0,\star}}{R_{\rm UFD}} +3.28037\right)+\frac{1}{4 R_{\rm UFD}}\right)/dt
\end{align}
Finally,
\begin{align}
\frac{d\mathcal{E}_\star}{dt}= G M_\star\frac{dR_{0,\star}}{dt}
\times \left\{\frac{3 \pi  M_\star}{64 R_{0,\star}^2}+\frac{M_{\rm UFD} R_{0,\star}}{4R_{\rm UFD}^3}\left(6  \log
   \frac{R_{\rm UFD}}{R_{0,\star}}+  (3 \log (4)-17)\right)\right\}
\end{align}

Note that this procedure has assumed that $R_{\rm UFD}$ stays fixed but the kinetic energy of the smooth component increases in order to reestablish hydrostatic equilibrium after the increase in $R_{0,\star}$. In practice, it is not clear that the entire system stays in hydrostatic equilbrium throughout. However, by including every term that contains $M_\star$, we have overestimated $\mathcal{E}_\star$, and hence underestimated the expansion of $R_{0,\star}$. Thus, our bound setting procedure is conservative. Finally, there is credence in this choice of $\mathcal{E}_\star$, after-all, the N-body simulations in Refs.~\cite{zhu2018primordial,Stegmann:2019wyz} agree very well with the analytical estimate of Ref.~\cite{Brandt:2016aco} that employs this $\mathcal{E}_\star$.

A similar procedure produces
\begin{align}
    \mathcal{E}_{\rm DM}= -G M_{\rm MACHO}^{\rm tot}\left( M_\star  \left(F_2\left(R_{0,\star},r_{\rm smooth}\right)\right)+M_{\pbh}^{\rm tot} F_3\left(R_{\pbh}\right)\right.\nonumber \\\left.+M_{\rm smooth} F_5\left(R_{\pbh},R_{\rm smooth}\right)+M_\star F_4\left(R_{0,\star},R_{\pbh}\right)\right.+\nonumber \\\left.+M_{\rm smooth} F_5\left(R_{\rm smooth},R_{\pbh}\right)\right)
\end{align}

\bibliography{biblio.bib}

\end{document}